\documentclass[a4paper,11pt]{article}
\usepackage[utf8]{inputenc}
\usepackage{jcappub}
\usepackage{mathtools}

\title{Cosmological Analysis using Panstarrs data: Hubble Constant and Direction Dependence}
\author[a]{Rahul Kumar Thakur,}
\author[b]{Meghendra Singh,} 
\author[c,1]{Shashikant Gupta,\note{Corresponding author.}}
\author[a]{Rahul Nigam}

\affiliation[a]{Birla Institute of Technology $\&$ Science, Pilani- Hyderabad Campus,\\Hyderabad, 500078 India}
\affiliation[b]{Delhi Metro Rail Corporation Limited \\ New Delhi, 110001, India}
\affiliation[c]{G D Goenka University,\\ Gurgaon, Haryana, 122103, India}

\emailAdd{thakurr58@gmail.com}
\emailAdd{meghendrasingh\_db@yahoo.co.in}
\emailAdd{shashikantgupta.astro@gmail.com}
\emailAdd{rahul.nigam@hyderabad.bits-pilani.ac.in}

\abstract{
Hubble tension and the search for preferred direction are two crucial unresolved issues in modern cosmology. Different measurements of the Hubble constant provide significantly different values, and this is known as the Hubble tension. The cosmological principle assumes that the universe is homogeneous and isotropic; however, deviations from the isotropy have often been observed. 

We apply the Bayesian tools and the Extreme Value theory dependent statistic to address the above issues. These techniques have been applied to the Panstarrs1 type Ia supernovae data. Our analysis for Hubble constant does not reject the Hubble tension. However, our value is smaller than that of the SHoES program and agrees with the CCHP value. Extreme value theory-based analysis indicates that the data does not show direction dependence. As a byproduct of our technique, we show that the errors in the data are non-Gaussian in nature.

}
\date{}

\begin{document}
\maketitle
\section{Introduction}
\label{sec:intro}
Recent observations of the Type Ia supernovae (SNe) have established that the expansion of the universe is accelerating \cite{riess1998observational,perlmutter1998discovery}. Along with the CMB observations, it established that the universe has flat geometry and is dominated by dark energy and dark matter. The $\Lambda CDM$ model \cite{astier2012observational} is one of the simplest cosmological models which explains all these observations and has been accepted as the standard model of cosmology. However, there are several challenges associated with the $\Lambda CDM$ cosmology. We discuss two of them below.

\subsection{Direction Dependence and Systematic Effects in Cosmological Data}
Cosmological principle (CP) is one of the major postulates of modern cosmology.  This assumption implies homogeneity and isotropy at cosmological scales, and the observed cosmological inhomogeneities are random fluctuations around a uniform homogeneous and isotropic background. The observations of CMB fluctuations provide excellent support to this principle \cite{bennet03}. 
Although CP plays a crucial role in modern cosmology, it is essential to determine whether the CP is consistent with the latest cosmological data. 
Previous studies, \cite{gupta2010direction} and \cite{gupta2014high} have investigated the direction dependence in the type Ia supernovae data using the techniques based on extreme value statistics. They also obtained information about non-Gaussian features in the data as a byproduct of the technique. Several other studies such as \cite{anton10} have found systematic issues with the high-redshift supernova data, including a direction dependence. 
A preferred direction in the CMB fluctuations may indicate a non-trivial topology\cite{moffatinh} of the universe. Others such as \cite{acute} and\cite{teg}  have explored such a preferred axis dominated by low multipoles which points towards $(l,b)\simeq(-100^\circ,60^\circ)$ and is termed as the axis of evil. 
    

\subsection{The Hubble constant and Hubble tension}
The Hubble constant ($H_0$) \cite{hubble1929} is one of the most important cosmological parameters in cosmology as it represents the expansion rate and sets the age of the Universe. The Hubble constant is also a valuable tool to calculate the critical density, $\rho_c = 3H_0/8\pi G$, required for the flat geometry of the Universe. Its numerical value helps to determine the various cosmological parameters, for instance, properties of galaxies and quasars, growth of large scale structures, etc.
Thus, knowing an accurate value of $H_0$ is of great importance in modern cosmology.  

The methods of measuring $H_0$ can be divided into two distinct classes: (a) direct measurement of $H_0$ by measuring the distance to SNe Ia or other astronomical probes. (b) indirect measurement through observations of the early universe and then applying the standard cosmological model. Coincidentally, both methods provide different values. The value obtained from the SHoES program \cite{riess20162} using the latest SNe Ia observations is $H_0 = 73.24\pm 1.74$ km/S/Mpc. which is 3.4 $\sigma$ 
away from the Planck value $H_0 = 67.8\pm 0.9 \pm 1.1$ provided by the CMB observations \cite{ade2014planck}. Another SNe Ia based observations from the CCHP team \cite{riess20162} is $H_0 = 69.8\pm 0.8 \pm 1.1$ which is around $3 \sigma$ away from the Planck value \cite{marra13, bennet14}. This discrepancy is termed as the Hubble tension. 
In principle, the distance measurement in the direct method may suffer from various systematic effects which could be responsible for the difference. However, the technological advancement in the last two decades has led to a significant improvement in the precision and accuracy of distance measurement. If we assume that both the methods are free of systematic effects than a non-standard cosmological model, such as a dark component in the early universe could be a possible solution to the Hubble tension.

In the present paper, we plan to use the recent observations of SNe Ia for testing the direction dependence systematics and measurement of the Hubble constant. Various sections of the paper are organized as follows: In section \ref{sec:data} we describe the SNe Ia data we have used for our analysis, and in \ref{sec:method} we briefly explain our methods. The results are presented in \ref{sec:result} and finally we conclude in \ref{sec:conclusion}. 

\section{Data}
\label{sec:data}
This paper uses 335 low-redshift SNe Ia from the Pan-STARRS1 (PS1) medium-deep survey \cite{rest2014cosmological} for cosmological analysis. The main goal of the medium-deep survey of PS1 was to detect and monitor a large number of SNe Ia to accurately measure the dark energy equation of the state parameter, $w$. Some exciting qualities of the PS1 system are: (i) An optical design with a 1.8 m diameter primary mirror that can deliver images with low distortion within a 3.3 deg field of view (ii) back-illuminated CCDs, each having 800x800 pixels. 
In the last two decades, there have been many SNe surveys with multiple passbands and dense time-sampling such as SDSS \cite{Lampeit}, ESSENCE \cite{narayan} and SNLS \cite{Betoule}, due to which sample sizes have increased. Systematic uncertainties related to the calibration, selection effects, correlated flows, extinction correction and LC modelling have been analysed in these surveys. However, some other issues such as the properties of host galaxies and inconsistencies of Sn colour with Milky Way like reddening law have also been found potential sources of the systematic errors \cite{scolnic} Apart from including the above systematic effects in the analysis, PS1 provides a large SNe sample with a single instrument. \\
Table $7$ and $8$ of \cite{rest2014cosmological} contain $113$ low redshift SNe and $222$ PS1 SNe respectively. Out of the total $335$, SNe Ia $333$ have been used in this paper, as we could not find the remaining two SNe positions. The data contains redshift in CMB frame and distance modulus along with the other variables.

\section{Methodology}
\label{sec:method}
\subsection{Estimation of Cosmological Parameters}
\label{sec:method-param}
The maximum likelihood method is often used to obtain the best-fit parameters for a given model. The likelihood, $P(D|H)$, is the probability of obtaining the data assuming that the model $H$ is correct. Mathematically  likelihood is defined in terms of $\chi^2$ as follows: 
\begin{equation}
P(D|H) \propto \exp{(-\chi^2/2)} \,  ,
\label{eq:likeli}
\end{equation}
In the context of SNe Ia distances in the flat $\Lambda$CDM cosmology, with density parameter $\Omega_M$ and Hubble constant $H_0$ as model parameters, the $\chi^2$ can be defined as 
\begin{equation}
\chi^2 = \sum_{i=1}^N \left( \frac{\mu^{obs}_i-\mu(z_i;\Omega_M,H_0)}{\sigma_i} \right)^2 \, ,
\label{eq:chisq}
\end{equation}
where $\mu_i$ is the distance modulus, $z_i$ is the redshift, and $\sigma_i$ is the measurement uncertainty in the $i^{th}$ data point. Best-fit value of the parameters are the ones which maximize the Likelihood or minimize $\chi^2$.

The above method does not provide the direct probability of the cosmological model, so we use the Bayesian approach as well. The posterior probability of the parameters can be calculated using the Bayes theorem
\begin{equation}
     P(H|D) \propto P(D|H)\times P(H) \, ,
    \label{eq:bayes}
\end{equation}
 where $P(H)$ is the prior probability of the model, which represents our state of knowledge about the model. Priors could be subjective and hence should be chosen carefully; very strict priors should be avoided whenever possible. The other advantage of the Bayesian approach is the marginalization over the nuisance parameters. For instance, the cosmological models often require the matter density ($\Omega_M$) and the expansion rate ($H_0$) of the universe as the model parameters. If we are interested in the Hubble constant, we marginalize over the density parameter $\Omega_M$ through the following equation:  
\begin{equation}
    P({H_{0}/ \mu) = \int P(\mu/ \Omega_{M},H_{0})P(\Omega_{M},H_{0})d\Omega_{M}}
    \label{eq:margin}
\end{equation}
 The uniform priors within the range $0\leq \Omega_M \leq1$ has been used in our analysis. 
\subsection{Direction Dependence and non-Gaussian Features in Data}
\label{sec:method-dir}
To understand the direction dependence, we use $\Delta$ statistic originally introduced in \cite{gupta2010direction,gupta2014high} For completeness, a brief account of the same is provided below. Throughout our analysis, flat $\Lambda$CDM cosmology has been assumed. The best-fit values of cosmological parameters corresponding to a given cosmological model are calculated by minimizing the $\chi^2$ defined in Eq~\ref{eq:chisq}. 
The data set is divided into subsets to construct our statistic. One can define a plane to divide the sky into two hemispheres (North and South, say). Depending on the position of SNe Ia, the plane divides the data into two subsets. For each SN Ia following quantity is calculated using the best-fit values of the cosmological parameter through Eq~\ref{eq:chisq} 
\begin{equation}
\chi_j = \left( \mu_{obs}^{j} - \mu_{th}^{j}(z_j;\Omega_M)]/\sigma_j \right) \, .
\label{eq:chi1}
\end{equation}
Here, we have assumed that all the SNe Ia are statistically uncorrelated. One can calculate the normalized sum of $\chi_j^2$ for each subset as:
\begin{equation}
    \chi_{North}^2 = \chi^2_{Sub}/N_{Sub} = \frac{1}{N_{Sub}}\sum_{j\subset Sub} {\chi_j}^2 \, .
    \label{eq:chisub}
\end{equation}
$\chi_R^2$ ($R$ indicates the subset $North/South$) defined in the above equation is a measure of deviation or statistical scatter of the subset from the original cosmological model. Assuming that the cosmological principle holds, the SN magnitude should not depend on the direction. Hence, SNe in different directions should be scattered similarly with reference to the best-fit cosmology. Similarly, for the other subset, the sum defined in Eq~\ref{eq:chisub} can be calculated (say $\chi_{South}^2$). The difference of $\chi^2$'s in the two hemispheres is now calculated as
\begin{equation}
    \Delta_{\chi^2} = |\chi^2_{North} - \chi^2_{South}|
\end{equation}
A large value of $\Delta_{\chi^2}$ indicates the mismatch in the best-fit values of cosmological parameters along the defined direction. We are interested in the maximum difference for which we rotate the plane and calculate the difference each time. The maximum absolute difference ($\Delta$) is now calculated as:  
\begin{equation}
\Delta = {\rm max} \{| \Delta \chi^2 |\}\,\, . 
\end{equation}
To calculate the distribution of $\Delta$ understanding of extreme value theory is helpful \cite{kendall1977advanced}. As shown in \cite{gupta2014high}, $\Delta$ follows a non-symmetric distribution known as the Gumbel distribution. 
\begin{equation}
    P(\Delta) = \frac{1}{s} \exp{\left(-\frac{\Delta-m}{s}\right)} \exp{\left(-\exp{  ( -\frac{\Delta-m}{s}} ) \right)} \, 
\end{equation}
The distribution is completely determined by the shape parameter ($s$) and the position parameter ($m$). We need the theoretical distribution of $\Delta$, which is obtained by simulating many sets of normally distributed $\chi_i$ on the SNe Ia positions and calculating $\Delta$ from each set. We also calculate a bootstrap distribution by shuffling the data values over the SNe Ia positions. 

\begin{center} 
\begin{table}
\centering
\begin{tabular}{ |c|c|c|} 
 \hline
 $\Omega_{0}$ & $H_0$ &  $\chi_{\nu}^2$ \\ 
\hline
 0.25 & 70.0 & 0.97 \\
\hline
\end{tabular}
\caption{Best-fit value of parameters for a flat $\Lambda CDM$ cosmology from Panstarss data by minimising $\chi^2$.}
\label{Table:chisq}
\end{table}
\end{center}
\begin{center} 
\begin{table}
\centering
\begin{tabular}{ |c|c|} 
 \hline
Prior &  $H_0$  \\
 \hline
  Uniform &  70.1$\pm 0.9$ \\
  \hline
  \end{tabular}
\caption{Best-fit values of Hubble constant ($H_0$) after Bayesian marginalization on $\Omega_M$ is applied.}
\label{Table:H0-fulldata}
\end{table}
\end{center}
\begin{center}
\begin{table}
\centering
\begin{tabular}{ |c|c|c|c|c|} 
 \hline
 $\Delta_{\chi}^2$ & longt & lat \\ 
\hline
 0.37 & 126.14 & 24.47 \\
  \hline
\end{tabular}
\caption{Direction for maximum $\Delta$  in the Panstarrs data.}
\label{Table:deltachisq}
\end{table}
\end{center}
\section{Results}
\label{sec:result}
We first obtain the best-fit values of the cosmological parameters for $\Lambda$CDM cosmology from the full Panstarr data. These values are $\Omega_M=0.25$ and $H_0=70.0$km/s/Mpc respectively with $\chi^{2}$ per dof $0.97$ (see table~\ref{Table:chisq}).
\subsection{Estimation of $H_0$ and Hubble Tension}
\begin{figure}
\centering
\includegraphics[width=10.0cm]{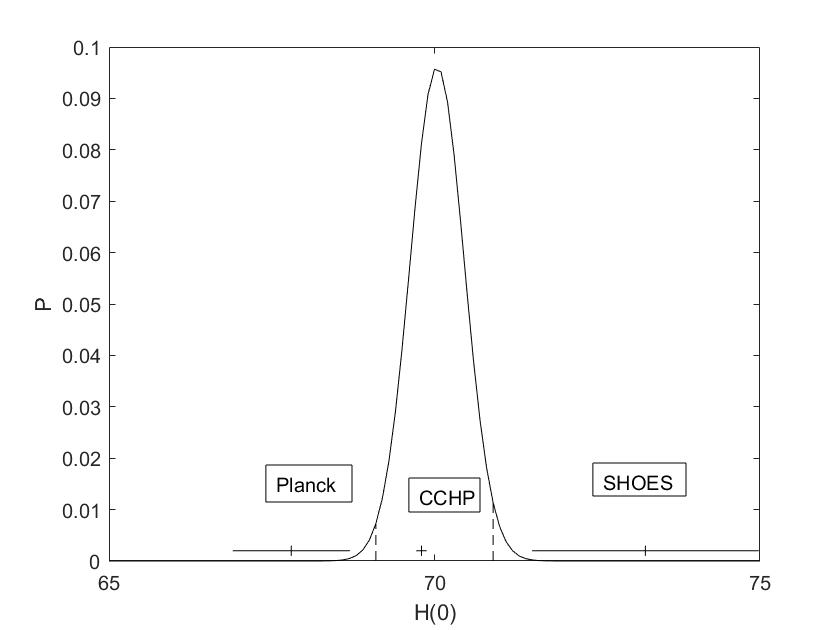}
\caption{Distribution of posterior probability of $H_{0}$ after marginalization for Panstarss Data.}
\label{fig:posterior}
\end{figure}

We now calculate the likelihood using Eq~\ref{eq:likeli} and then use the Bayesian marginalization to obtain the posterior probability for $H_0$. Uniform prior for $\Omega_M$ has been considered over a reasonable range. Throughout our calculation, we have considered a flat $\Lambda$CDM cosmology. The best-fit value of $H_0$, after marginalization over the density parameter, is presented in table~\ref{Table:H0-fulldata}. The posterior probability of $H_0$ is plotted in fig~\ref{fig:posterior}. We have also plotted the Planck \cite{ade2014planck} , SHoES \cite{riess20162} and CCHP \cite{freedman2019carnegie} results in the same graph. Our result matches well with the CCHP value. Planck and SHoES values appear on opposite sides of the posterior pdf. 
\subsection{Direction Dependence}
Now we apply the $\Delta$ statistics to the data. A plane is defined, which divides the sky into two hemispheres, and thus we obtain two subsets of the full data set. $\chi_R^2$ for each subset of data is calculated using the best-fit value of cosmological parameters (presented in table~\ref{Table:chisq}). The difference of the $\chi_R^2$ in both the subsets is obtained. The plane is now rotated to obtain the maximum value of this difference. The corresponding direction in which maximum is realized is presented in galactic coordinates in table~\ref{Table:deltachisq}. One can compare this direction with the axis of evil \cite{acute}. To understand the importance of this direction, we re-examine the best fit for the two subsets of the data. The number of SNe in both subsets are 87 and 246, respectively. The best-fit values of cosmological parameters for each of the subset are presented in table~\ref{Table:chisq-southern}. The values are quite different, especially $\Omega_M$ is quite small for one of the subsets. 

In order to estimate the significance of the direction of maximum $\Delta$ we need to obtain the distribution of $\Delta$ by shuffling the positions of SNe in the data (bootstrap distribution) as mentioned in \ref{sec:method-dir}. The distribution is plotted in fig~\ref{fig:distribution}. The original value of $\Delta$ from the data lies within $1 \sigma$ region, indicating that the direction is probably just by chance. 
\subsection{Non-Gaussian Errors in Data}
\begin{figure}
\centering
\includegraphics[width=10.0cm]{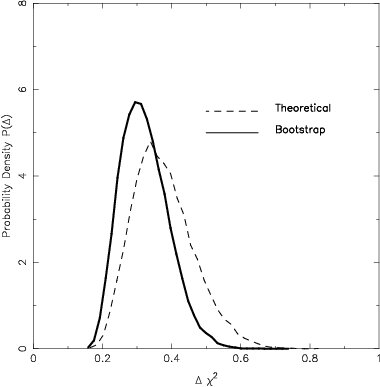}
\caption{Bootstrap and theoretical probability distributions for simulated data have been shown for comparison. Positions for $333$ SNe on the sky were generated randomly.}
\label{fig:Simulated}
\end{figure}

\begin{figure}
\centering
\includegraphics[width=10.0cm]{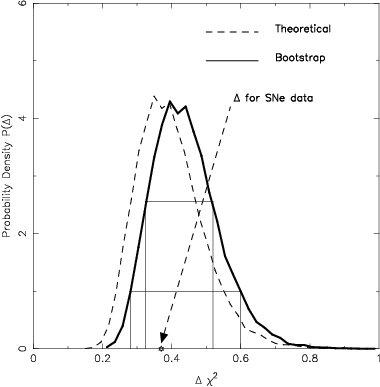}
\caption{The bootstrap probability distribution along with theoretical distribution for $\Delta$. This graph should be compared with figure~\ref{fig:Simulated} in which the theoretical distribution appears slightly on the right of bootstrap distribution. The opposite trend in the graph indicates non-Gaussian errors in the PS1 data.}
\label{fig:distribution}
\end{figure}

\begin{figure}
\centering
\includegraphics[width=15.0cm]{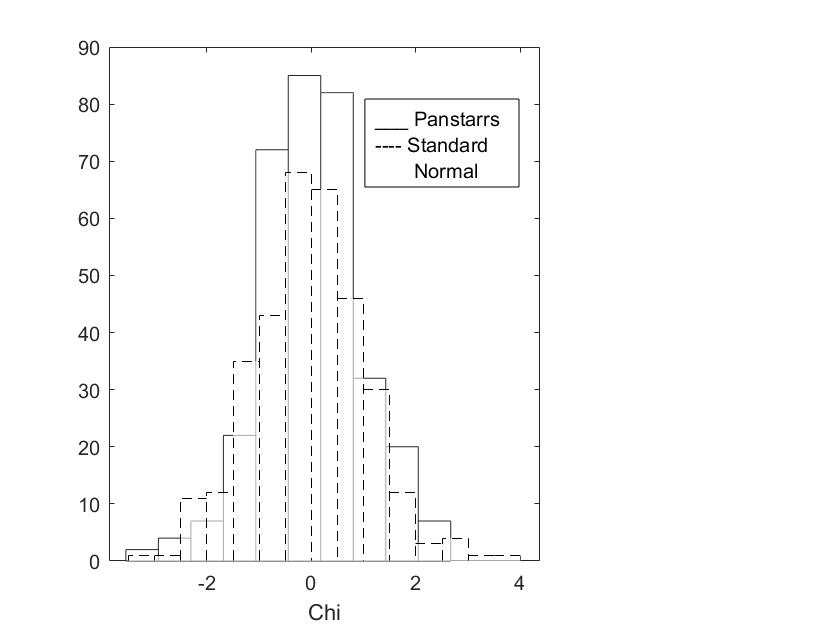}
\caption{Histogram of$\chi_{i}'s$ is compared with that of standard normal distribution.}
\label{fig:hist}
\end{figure}

For further investigation, in fig~\ref{fig:distribution}, we also plot the theoretical distribution of $\Delta$ by generating the Gaussian random variates as discussed in \ref{sec:method-dir}. For reference, the simulations of bootstrap and theoretical distributions have been plotted by generating 333 random positions in the sky in fig~\ref{fig:Simulated}. As discussed in \cite{gupta2014high}, theoretical $\chi$'s are unbounded; however, the bootstrap distribution will have an upper bound. Due to this reason, the bootstrap distribution in fig~\ref{fig:Simulated} appears slightly on the left compared to the theoretical distribution. This feature is also expected in the original distributions. However, the opposite has been observed in fig~\ref{fig:distribution}. If the errors in the data, $\sigma_j$, obey Gaussian distribution, then $\chi_j$'s in eq~\ref{eq:chi1} should follow the standard normal distribution. Thus the theoretical distribution has been produced by generating the standard normal distribution in place of actual $\chi_i$'s from the data. Mismatch with the simulations indicates that the data errors do not strictly obey the Gaussian distribution. 
To explore this further, we plot a histogram of $\chi_i$'s and compare it with that of the standard normal distribution in fig~\ref{fig:hist}. As both the histograms look different, it confirms the previous observation that the errors contain the non-Gaussian part.  

\begin{center}
\begin{table}
\centering
\begin{tabular}{ |c|c|c|c|c|} 
 \hline
Subset & $\Omega_{0}$ & $H_0$ &  $\chi_{\nu}^2$  \\ 
\hline
 Southern hemisphere  & 0.14 & 71 &   1.24\\
  \hline
 Northern hemisphere  & 0.28 & 69.7 &   0.88\\
 \hline
\end{tabular}
\caption{Best-fit values of Cosmological parameters from the two subsets of data in the direction of maximum $\Delta$. The name Southern and Northern hemispheres are arbitrary.}
\label{Table:chisq-southern}
\end{table}
\end{center}
\begin{figure}
\centering
\includegraphics[width=10.0cm]{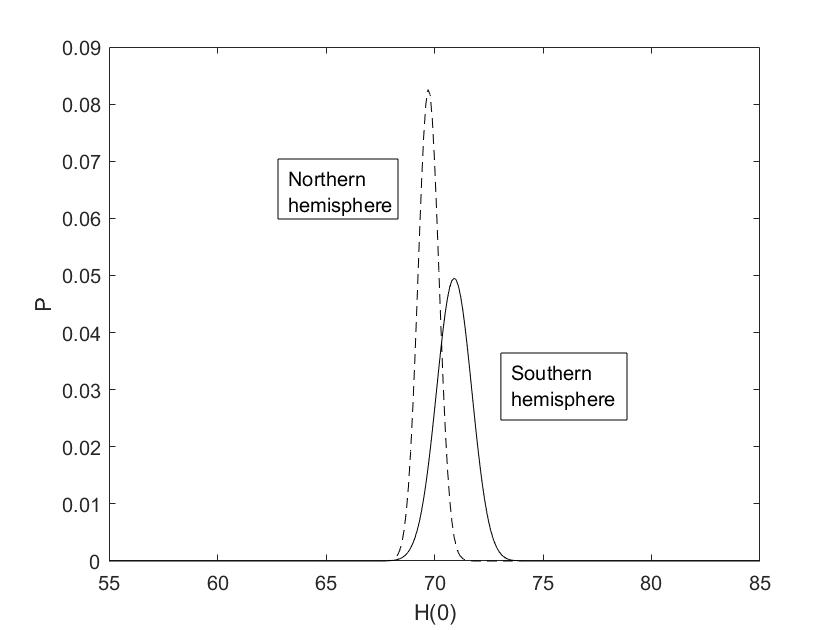}
\caption{Probability Distribution of $H_{0}$ values after marginalization for northern and southern hemisphere at 2 $\sigma$.}
\label{fig:combine hemisphere}
\end{figure}

\section{Conclusion}
\label{sec:conclusion}
We have estimated the Cosmological parameters using the Panstarrs data for the flat $\Lambda$CDM cosmology. The matter density and the Hubble constant obtained are within the usual range obtained from the other probes \cite{domin,taube}. We have also applied the Bayesian approach to estimate the Hubble constant by marginalizing over the density parameter. Our result matches well with the CCHP result \cite{freedman2019carnegie}. However, there is an apparent mismatch with both the Planck and SHoES values. Our value lies in the middle of the two. One may consider the presence of some systematic effects in the SHoES data, especially with the calibration of MW Cehpeids as there is growing evidence in favour of the CCHP value \cite{hoyt2019carnegie,jang2021carnegie,ferrarese2000hubble}. It may relax the Hubble tension to a smaller level. However, the Hubble tension still exists as the Planck value is entirely outside the posterior probability curve. 

We also investigate the direction dependence in the data. Our results using the $\Delta$ statistic show a mild direction dependence which could be a coincidence or a chance. On the other hand, we show that the uncertainties in the data do not strictly obey the Gaussian distribution, which possibly reflect calibration issues in the Panstarrs data. 

\section*{Acknowledgement}
Shashikant Gupta thanks SERB (India) for financial assistance (EMR/2017/003714).

\end{document}